\documentclass[aps,prl,superscriptaddress,onecolumn,showpacs,amsmath,amssymb]{revtex4}

\usepackage{color}
\usepackage{graphicx}
\usepackage{bm}
\definecolor{Blue}{rgb}{0.00, 0.00, 1.00}
\definecolor{Red}{rgb}{1.00, 0.00, 0.00}

\begin{document}

\title{Gate-controlled mid-infrared light bending with aperiodic graphene nanoribbons array}
\author{Eduardo Carrasco$^{\dagger}$ }
\affiliation{Adaptive MicroNano Wave Systems, Ecole Polytechnique Federale de Lausanne (EPFL), 1015 Lausanne, Switzerland}

\author{Michele Tamagnone$^{\dagger}$ }
\affiliation{Adaptive MicroNano Wave Systems, Ecole Polytechnique Federale de Lausanne (EPFL), 1015 Lausanne, Switzerland}
\affiliation{Laboratory of Electromagnetics and Acoustics (LEMA), Ecole Polytechnique Federale de Lausanne (EPFL), 1015 Lausanne, Switzerland}

\author{Juan R. Mosig}
\affiliation{Laboratory of Electromagnetics and Acoustics (LEMA), Ecole Polytechnique Federale de Lausanne (EPFL), 1015 Lausanne, Switzerland}

\author{Tony Low$^{*}$ }
\affiliation{Department of Physics \& Electrical Engineering, Columbia University, New York, NY 10027, USA}
\affiliation{Department of Electrical \& Computer Engineering, University of Minnesota, Minneapolis, MN 55455, USA}

\author{Julien Perruisseau-Carrier}
\affiliation{Adaptive MicroNano Wave Systems, Ecole Polytechnique Federale de Lausanne (EPFL), 1015 Lausanne, Switzerland}

\date{\today}
\begin{abstract}
\textbf{
Graphene plasmonic nanostructures enable subwavelength confinement of electromagnetic energy from the mid-infrared down to the terahertz frequencies.  
By exploiting the spectrally varying light scattering phase at vicinity of the resonant frequency of the plasmonic nanostructure, it is possible to control the angle of reflection of an incoming light beam.
We demonstrate, through full-wave electromagnetic simulations based on Maxwell equations, the electrical control of the angle of reflection of a mid-infrared light beam by using an aperiodic array of graphene nanoribbons, whose widths are engineered to produce a spatially varying reflection phase profile that allows for the construction of a far-field collimated beam towards a predefined direction. 
}
\newline
$^\dagger$ These authors contributed equally\\
$^*$ Corresponding email: tlow@umn.edu\\
\end{abstract}
\maketitle

\emph{Introduction---} Graphene plasmonics has emerged as a very promising candidate for terahertz to mid-infrared applications \cite{Low14}, the frequency range where it supports plasmonic propagation. \cite{Ju11, Koppens11, Grigorenko12, Yan12, Yan13, Chen12, Fei12, Jablan09, Low14}. Today the terahertz to mid-infrared spectrum, ranging from $10\,$cm$^{-1}$ to $4000\,$cm$^{-1}$, is finding a wide variety of applications in information and communication, homeland security, military, medical sciences, chemical and biological sensing, spectroscopy, among many others \cite{Tonouchi07, Ferguson02, Soref10}. The two-dimensionality of graphene and its semi-metallic nature allow for electrical tunability not possible with conventional metals by simply biasing electrostatically the graphene device. In addition, graphene is also an excellent conductor of electricity, with highest attained carrier mobility reaching $1,000,000\,$cm$^2$/Vs in suspended samples \cite{Du08} and $100,000\,$cm$^2$/Vs for ultra-flat graphene on boron nitride \cite{Dean10}. These attributes, in addition to its stability and compatibility with standard silicon processing technologies, have also raised interest in a myriad of graphene-based passive and active photonic \cite{Tamagnone14, Bonaccorso12, Bao12} and terahertz devices \cite{Yan12, Ju11, Lee12, Otsuji12, Vicarelli12, Chen13, Padooru13}.

Electromagnetic surfaces able to dynamically control the reflection angle of an incident beam have been studied for decades in the microwave community \cite{Hum13, Huang08}, with applications in satellite communications, terrestrial and deep-space communication links. Non-conventional reflecting surfaces for optical frequencies have also been proposed recently, using metal elements having certain fixed configurations and operating in the plasmonic regime \cite{Yu11, Ni12}. However, the high carrier concentration in metals prohibits dynamic control of the reflected beams via these surface elements. In this letter, we show how the reflection angle of a mid-infrared incident beam can be electrically controlled with an aperiodic array of graphene nanoribbons, by utilizing its intrinsic plasmonic resonance behavior. We demonstrate that the proposed effect is experimentally realizable by performing numerical Maxwell simulations of the device with experimentally feasible parameters.\\


\emph{Basic Concepts---}
Consider an aperiodic array of graphene nanoribbons of varying widths lying at the interface of two half spaces as illustrated in Fig.\,1a, with $p$-polarized light at incidence angle $\theta_i$, which is reflected and transmitted at angle $\theta_r$ and $\theta_t$ respectively. Each nanoribbon constitutes a plasmonic resonator, which can effectively produce a scattering phase $\phi$ between $0$ and $-\pi$ depending on the frequency of free-space light $\omega$ with respect to the plasmon resonance frequency $\omega_0$. In plasmonic metasurfaces consisting of nanoribbons of varying widths, the scattering phase in general can also vary across the interface, namely as $\phi(x_j)$, where $x_j=j\Delta x$ and $\Delta x$ being the distance between the centers of adjacent ribbons, which is considered constant in this contribution. In the limit where ray optics is applicable, i.e. $\lambda\gg \Delta x$, where $\lambda$ is the free-space wavelength, the generalized Snell's law dictates that \cite{Yu11},
\begin{eqnarray}
\nonumber
\mbox{sin}(\theta_t)\sqrt{\epsilon_2}-\mbox{sin}(\theta_i)\sqrt{\epsilon_1}
=\frac{\lambda}{2\pi}\frac{d\phi}{dx}\\
\mbox{sin}(\theta_r)-\mbox{sin}(\theta_i)=\frac{\lambda}{2\pi \sqrt{\epsilon_1}}\frac{d\phi}{dx}
\label{snell}
\end{eqnarray}
Eq.\,\ref{snell} implies that the reflected or transmitted beam can be effectively bent such that $\theta_r\neq \theta_{i,t}$ if a constant spatial gradient in the scattering phase is imposed (i.e. $d\phi/dx=\mbox{constant}$). Simple estimates from Eq.\,\ref{snell} suggest that it is possible to bend normal incident ($\theta_i=0$) mid-infrared light far from broadside (i.e. $\theta_r\neq 0$). Assuming $\lambda=10\,\mu$m (e.g from a CO$_2$ laser), $\Delta x=100\,$nm (typical ribbon width where plasmon resonance resides in the mid-infrared), one obtains $\Delta\phi\ll \pi$, suggesting that it is indeed possible to induce a gradual spatial variation in $\phi$ across the interface.

The scattering phase due to a graphene plasmonic resonator, and its design space, can be examined more quantitatively as follows.  We consider graphene on a SiO$_2$ substrate and with an electrolyte superstrate, which can also serve as a top gate for inducing high doping in graphene \cite{Efetov10}. We solve the Maxwell equations for the reflection coefficient of a $p$-polarized light, $r_p(q,\omega)$, where graphene is modeled by its dynamic local conductivity $\sigma(\omega)$ obtained from the random phase approximation  \cite{Wunsch06, Hwang07}. 
The scattering phase $\phi$ then follows from $\phi(q,\omega)= \mbox{arg}[r_p(q,\omega)]$. The nanoribbon resonator frequency $\omega_0$ can be estimated from the scattering coefficients for a continuous monolayer, where the in-plane wave-vector $q$ is related to the width $W$ of the nanoribbon via $q=3\pi/4W$, after accounting for the anomalous reflection phase off the edges \cite{Nikitin14}:
\begin{eqnarray}
\omega_0 = \sqrt{\frac{3e^2\mu }{8W\hbar^2\epsilon_{env}}}
\label{omega0}
\end{eqnarray}
where $\epsilon_{env}=\tfrac{1}{2}(\epsilon_1+\epsilon_2)$. The Lorentz oscillator model provides then a simple expression for the scattering phase:
\begin{eqnarray}
\phi(W,\omega)\approx \mbox{tan}^{-1}\left(  \frac{\omega}{\tau_0}\left(   \omega^2 - \frac{3e^2\mu }{8W\hbar^2\epsilon_{env}}      \right)^{-1} \right)
\label{simple}
\end{eqnarray}
In this simple model calculation, we simply take $\epsilon_2=3.9$ and $\epsilon_1=6$. Assuming $\lambda=10\,\mu$m, Fig.\,1b-c depicts  $\phi(W,\omega)$ for varying $W$, with graphene at different chemical potentials $\mu$, and electronic lifetimes $\tau_0$. The phase $\phi$ changes most rapidly with $W$ when the ribbon is at resonance. For a given ribbon plasmon resonance frequency, increasing the doping would allow for the same resonance frequency at a larger $W$ as depicted in  Fig.\,1b. On the other hand, decreasing $\tau_0$ dampens the plasmon resonance, causing a smoother variation in $\phi$ with $W$ as shown in Fig.\,1c. Eq.\,\ref{simple} therefore provides a simple intuitive understanding of the scattering phase of a graphene plasmonic resonator. \\

\emph{Device Simulation---}
In principle, the tunable scattering phase of a graphene plasmonic resonator allows the design of surface elements which controls the angle of reflection or transmission of an incoming beam. In this work, we consider the former i.e. a reflectarray \cite{Huang08}. Fig.\,2a-b shows a detailed schematic of the proposed device, designed for a working frequency of $27\,$THz ($900\,$cm$^{-1}$). The graphene nanoribbons array is between an electrolyte gating superstrate of $200\,$nm and a $1.2\,\mu$m SiO$_2$ dielectric substrate with a metal layer underneath, which serves as a reflector, reflecting most of the incoming light. 
The full dielectric function of SiO$_2$ is used in the simulation \cite{Palik98}. For the electrolyte superstrate, we assumed a dielectric constant of $6$. Hence, the effective dielectric constant of the graphene's environment, which ultimately determines its plasmonic response, is approximately $5$, similar to Ref. \cite{Ju11}.
The aperiodic array of nanoribbons has a designed inter-ribbon separation $p$ taken to be $140\,$nm, and the width of each nanoribbons is to be chosen so that the spatial variation in $\phi_r$ satisfies $d\phi_r/dx=\mbox{const.}$, as discussed previously. As explained later, the final reflection phase $\phi_r$ of the reflecting cell is not given simply by the response $\phi$ of the ribbon, since the contribution of the ground plane must also be taken into account. We consider a Gaussian beam illuminating the array at an incident angle $45^o$ with respect to the normal of the $xz$ plane as illustrated. Fig.\,2c depicts the top view of the nanoribbons array, including the elliptical projection of the impinging beam on the $xz$ plane, where $b_x=31.36\,\mu$m and $b_y=22.18\,\mu$m. 

We use CST Microwave Studio to numerically compute the reflection coefficient and $\phi_r$ for a nanoribbon of particular width. Using the Floquet’s theory for periodic arrays, mutual coupling between neighboring nanoribbons is accounted for. Graphene is modeled by its dynamic local conductivity $\sigma(\omega)$, assuming typical electronic lifetime of $\tau_0=0.1\,$ps. Fig.\,3a shows the calculated $\phi_r$ of the reflection coefficient, as a function of the chemical potential $\mu$ and width $W$ of the nanoribbon. For a chemical potential $\mu=1.0\,$eV, $\phi_r$ varies between $0^o$ and $-340^o$ by adjusting the width of the ribbons between $40\,$nm and $140\,$nm. As $\mu$ decreases, the range of $\phi_r$ decreases, which approaches a constant $\phi_r=-340^o$ at $\mu=0.3\,$eV. 

We design the reflectarray device to perform a binary operation: consisting of producing a reflected far-field beam in the broadside direction ($0^o$) and a reflected far-field beam in the specular direction ($45^o$). The  spatial phase profile required to achieve these reflection angles can be determined based on Eq.\,\ref{snell}, and are shown in Fig.\,3b for these two cases. For specular beam, a zero phase difference between the nanoribbons would suffice, regardless of the absolute value of that phase. Otherwise, the sequence of ribbons' widths has to be chosen such that it provides the neccessary spatial phase profile to produce a reflected far-field beam at $\mu=1.0\,$eV. As mentioned above, with ribbons between $40\,$nm and $140\,$nm, $\phi_r$ can be made to vary between $0^o$ and $-340^o$. The missing phase values (from $-340^o$ to $-360^o$) were found not to have noticeable impact in the far-field generation. The widths of the nanoribbons along the array are displayed in Fig.\,3c. In order to produce a reflected far-field beam in the specular direction, a constant $\phi_r$ is required along all the elements of the array. This can be achieved by decreasing $\mu$ to $0.3\,$eV as shown. In other words, the binary operation of our reflectarray device can be achieved by electrically controlling the doping $\mu$. 

In Fig.\,4a and b, we compute the far-field produced by our reflectarray device, for broadside ($\mu=1.0\,$eV) and specular radiation ($\mu=0.3\,$eV) respectively.  The possibility of bending the light beam in an aperiodic array of graphene nanoribbons is clearly demonstrated by these simulations. \\


\emph{Discussion---}
A constant phase gradient is needed for producing a collimated beam. At intermediate dopings, where the ribbon array phases are not designed with constant phase gradient, the produced far field beam can be highly distorted. The design of smooth beam steering is possible, but would require a more complicated gating scheme that addresses the doping of individual ribbons separately as typically done in microwave reflectarrays \cite{Kamoda11} and more recently also for terahertz graphene-based reflectarrays \cite{Carrasco13}. The simplicity in the reflectarray array scheme proposed in this work has the obvious appeal of providing a design much easier to be implemented experimentally.  

A simple circuit model for each nanoribbon cell, as shown in Fig.\,2d, is useful for better understanding of the proposed device. The substrate and superstrate can be modelled with two transmission line segments having propagation constants and characteristic impedance equal to the two media which they model. The equivalent of the ground plane is simply a short circuit, while the graphene ribbons can be represented by an equivalent $Z_g$ impedance in parallel, that models the surface currents induced in the ribbons by the tangential electric field. An approximate closed form expression for $Z_g$ can be found by modeling resonant ribbons as Lorentz oscillators as explained previously. 
This can be represented by a simple RLC series circuit, where $R$ and $L$ are simply found from the real and imaginary parts of the surface conductivity of graphene $\sigma$, while $C$ represents the quasi-static electric fields associated with the plasmons, and can be obtained by enforcing the resonant frequency of the RLC circuit ($1/\sqrt{LC}$) to be equal to the ribbons resonant frequency $\omega_0$:
\begin{equation}
L=\frac{\pi\hbar^2}{e^2\mu} \qquad  R=\tau^{-1}L \qquad C = \frac{1}{\omega_0^2L}
\label{rlc}
\end{equation}
where
\begin{equation}
\omega_0 \approx \sqrt[]{\frac{3e^2\mu}{8W\hbar^2\epsilon_{env}}}
\end{equation}
The final equivalent impedance of graphene nanoribbons is then,
\begin{equation}
Z_g = R + j\omega L + \frac{1}{j\omega C} = \frac{\pi\hbar^2}{e^2\mu}\left(\frac{\omega_0^2+j\omega\tau^{-1}-\omega^2}{j\omega}\right)
\label{zg}
\end{equation}
This circuit model provides for an estimated value for the reflection phase $\phi_r$ as function of $W$ and $\mu$. Fig.\,3d shows the calculated reflection phase $\phi_r$ which agrees qualitatively with the numerical simulations presented in Fig.\,3a. The full wave simulations account for higher order phenomena not captured by this simple model.  Importantly, while in the Lorentz oscillator the phase range is always less than $\pi$ for the nanoribbons alone (i.e. Eq.\,\ref{simple}), the presence of the substrate and of the ground plane allows here a much wider phase range, which is a key point for the device presented in our work. 

Our device has an average element loss  of $1.32\,$dB for the $\mu=0.3\,$eV configuration, and $1.77\,$dB for $\mu=1\,$eV. Hence, most of the incoming light is reflected. The loss of the device is directly related to the assumed electronic lifetime $\tau$, as evident from Eq.\,\ref{rlc}, and can be improved with better graphene quality or by using a better substrate such as boron nitride \cite{Woessner14}. \\

\emph{Conclusion---}
In this work, we show how the reflection angle of a mid-infrared beam can be dynamically controlled by employing tunable graphene plasmons in an aperiodic array of graphene nanoribbons according to the generalized laws of reflection. The attractive feature of our device proposal, the use of a simple gating scheme, renders it more experimentally feasible and practical. Numerical Maxwell device simulations convincingly demonstrate the dynamic bending of mid-infrared beams, assuming experimentally accessible physical parameters.     \\

\emph{Acknowledgements---}
This work was supported by the European Union (Marie Curie IEF 300934 RASTREO project), the Hasler Foundation (Project 11149) and the Swiss National Science Foundation (Project 133583), and the University of Minnesota. E. Carrasco, M. Tamagnone, T. Low and J. R. Mosig would like to dedicate this paper to the memory of J. Perruisseau-Carrier, who passed away during the preparation of this work.


\newpage

\textbf{Figure Captions}\\

\textbf{Figure 1:}
(a) Schematic illustrating the dynamic control of the angle of reflected and transmitted light by introducing gradient of scattering phases at the interface using graphene nanoribbon resonators with different widths. 
(b-c) Calculated scattering phase $\phi$ for varying ribbons' width for $p$-polarized light with free-space wavelength $\lambda=10\,\mu$m, e.g from a CO$_2$ laser. Calculations are done for different chemical potentials $\mu$ and electronic lifetimes $\tau_0$ as shown. Symbols represent calculation from Maxwell equation, while solid lines are from simple model in Eq.\,\ref{simple}.\\

\textbf{Figure 2:}
(a) Schematic of the gate-controlled reflectarray device based on aperiodic array of graphene nanoribbons. (b) Top view of the graphene nanoribbons array, including the elliptical projection of the impinging beam.  (c) Lateral view of the device, highlighting the superstrate, substrate, ribbons, and metal reflector. (d) Equivalent circuit for one element of the proposed array.\\

\textbf{Figure 3:}
(a) Phase $\phi$ of the reflection coefficient as a function of chemical potential $\mu$ for different widths $W$ of a nanoribbon, ranging from $40\,$nm to $140\,$nm. Mutual coupling between neighboring ribbons is taken into account.  
(b) Spatial phase profile required at each nanoribbon along the $x$-axis in order to produce a far-field beam in the broadside direction ($0^o$) and a reflected far-field beam in the specular direction ($45^o$). 
(c) Width of the nanoribbons across the array ($x$-axis) of the reflectarray device. 
(d) Phase $\phi$ of the reflection coefficient as a function of chemical potential $\mu$ for different widths $W$ of a nanoribbon using the equivalent circuit model described in Fig.\,2d.
\\

\textbf{Figure 4:}
Far-field radiation pattern produced by the reflective array for the two bias condition; (a) Beam towards broadside ($0^o$) at $\mu=1.0\,$eV and (b)  specular direction ($45^o$) at $\mu=0.3\,$eV.


\newpage

\scalebox{0.6}[0.6]{\includegraphics*[viewport=50 0 650 650]{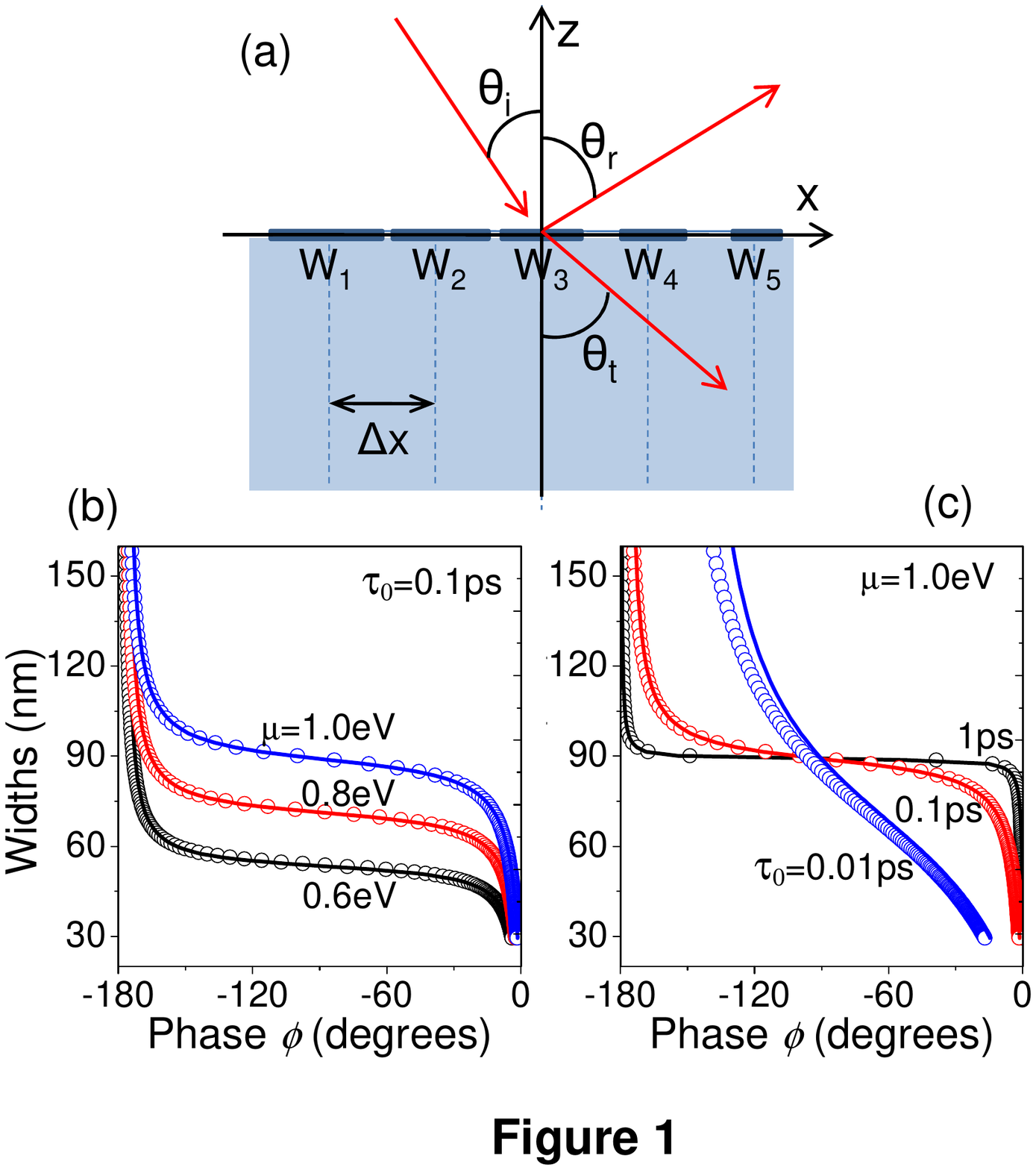}}
\newpage

\scalebox{0.6}[0.6]{\includegraphics*[viewport=50 0 700 650]{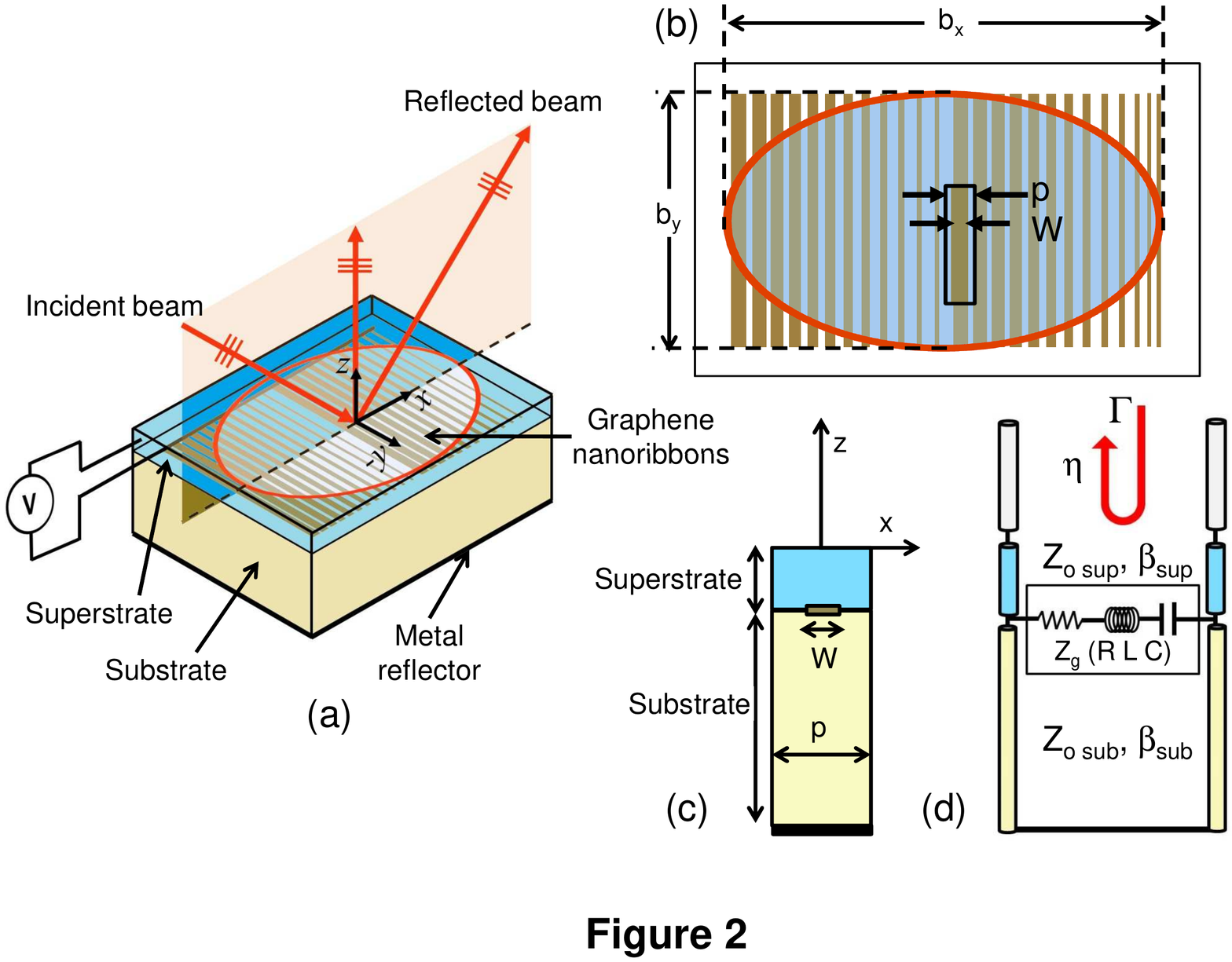}}
\newpage

\scalebox{0.6}[0.6]{\includegraphics*[viewport=50 0 700 650]{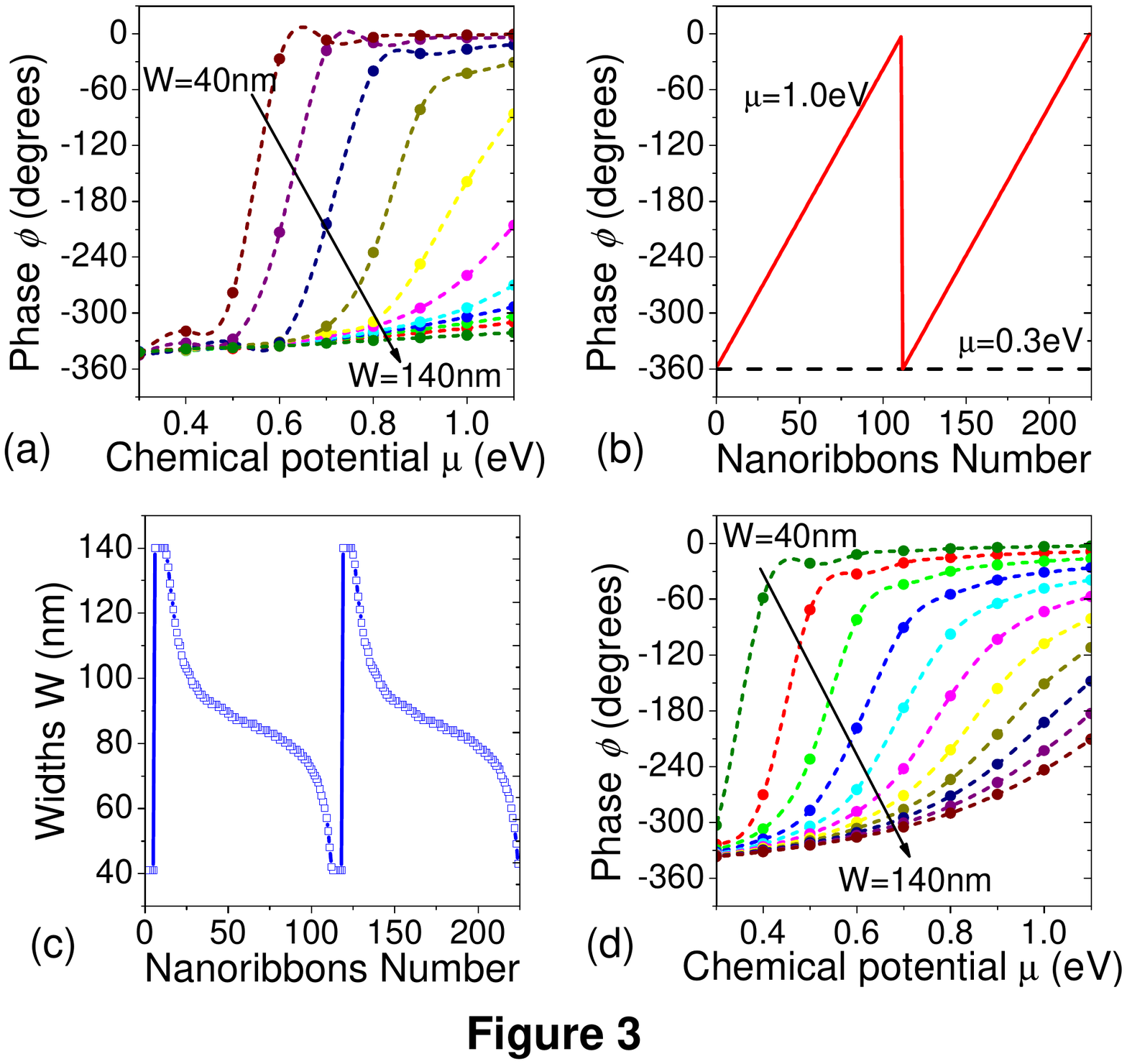}}
\newpage

\scalebox{0.6}[0.6]{\includegraphics*[viewport=50 0 700 650]{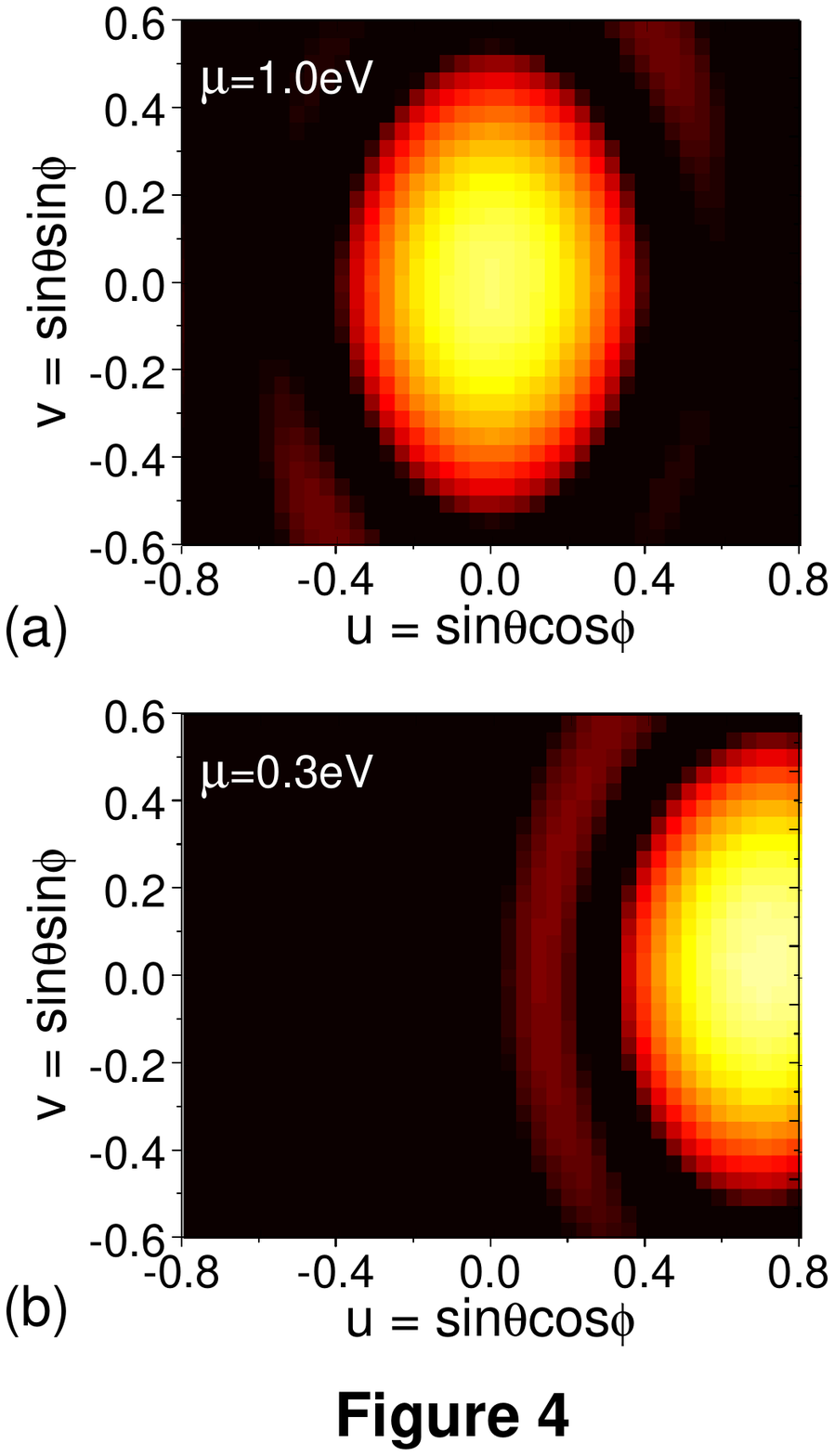}}


\newpage


\begin{thebibliography}{widest entry} 
\bibitem{Low14} Low, Tony, and Phaedon Avouris. ``Graphene Plasmonics for Terahertz to Mid-Infrared Applications." ACS nano 8, no. 2 (2014): 1086-1101.

\bibitem{Tamagnone14} Tamagnone, Michele, Arya Fallahi, Juan R. Mosig, and Julien Perruisseau-Carrier. ``Fundamental limits and near-optimal design of graphene modulators and non-reciprocal devices." Nature Photonics (2014).

\bibitem{Ju11} Ju, Long, Baisong Geng, Jason Horng, Caglar Girit, Michael Martin, Zhao Hao, Hans A. Bechtel et al. ``Graphene plasmonics for tunable terahertz metamaterials." Nature nanotechnology 6, no. 10 (2011): 630-634.

\bibitem{Koppens11} Koppens, Frank HL, Darrick E. Chang, and F. Javier Garcia de Abajo. ``Graphene plasmonics: a platform for strong light–matter interactions." Nano letters 11, no. 8 (2011): 3370-3377.

\bibitem{Grigorenko12} Grigorenko, A. N., Marco Polini, and K. S. Novoselov. ``Graphene plasmonics." Nature photonics 6, no. 11 (2012): 749-758.

\bibitem{Yan13} Yan, Hugen, Tony Low, Wenjuan Zhu, Yanqing Wu, Marcus Freitag, Xuesong Li, Francisco Guinea, Phaedon Avouris, and Fengnian Xia. ``Damping pathways of mid-infrared plasmons in graphene nanostructures." Nature Photonics 7, no. 5 (2013): 394-399.

\bibitem{Yan12} Yan, Hugen, Xuesong Li, Bhupesh Chandra, George Tulevski, Yanqing Wu, Marcus Freitag, Wenjuan Zhu, Phaedon Avouris, and Fengnian Xia. ``Tunable infrared plasmonic devices using graphene/insulator stacks." Nature Nanotechnology 7, no. 5 (2012): 330-334.

\bibitem{Chen12} Chen, Jianing, Michela Badioli, Pablo Alonso-González, Sukosin Thongrattanasiri, Florian Huth, Johann Osmond, Marko Spasenović et al. ``Optical nano-imaging of gate-tunable graphene plasmons." Nature 487, no. 7405 (2012): 77-81.

\bibitem{Fei12} Fei, Zhe, A. S. Rodin, G. O. Andreev, W. Bao, A. S. McLeod, M. Wagner, L. M. Zhang et al. ``Gate-tuning of graphene plasmons revealed by infrared nano-imaging." Nature (2012).

\bibitem{Jablan09} Jablan, Marinko, Hrvoje Buljan, and Marin Soljačić. ``Plasmonics in graphene at infrared frequencies." Physical review B 80, no. 24 (2009): 245435.

\bibitem{Tonouchi07} Tonouchi, Masayoshi. ``Cutting-edge terahertz technology." Nature photonics 1, no. 2 (2007): 97-105.

\bibitem{Ferguson02} Ferguson, Bradley, and Xi-Cheng Zhang. ``Materials for terahertz science and technology." Nature materials 1, no. 1 (2002): 26-33.

\bibitem{Soref10} Soref, Richard. ``Mid-infrared photonics in silicon and germanium." Nature Photonics 4, no. 8 (2010): 495-497.

\bibitem{Du08} Du, Xu, Ivan Skachko, Anthony Barker, and Eva Y. Andrei. ``Approaching ballistic transport in suspended graphene." Nature nanotechnology 3, no. 8 (2008): 491-495.

\bibitem{Dean10} Dean, C. R., A. F. Young, I. Meric, C. Lee, L. Wang, S. Sorgenfrei, K. Watanabe et al. ``Boron nitride substrates for high-quality graphene electronics." Nature nanotechnology 5, no. 10 (2010): 722-726.


\bibitem{Huang08} Huang, J., Encinar, J. A., ``Reflectarray Antennas", IEEE Press, USA, 2008. 

\bibitem{Chen13} P. Y. Chen, A. Alu, ``Terahertz Metamaterial Devices Based on Graphene Nanostructures", IEEE Trans. Terahertz Science and Tech., IEEE 2013. 

\bibitem{Padooru13} Y. R. Padooru, A. B. Yakolev, C. S. R. Kaipa, G. W. Hanson, F. Medina, F. Mesa, ``Dual capacitive-inductive nature of periodic graphene patches: transmission characteristics at low terahertz frequencies", Physical Review B87, 115401 (2013).

\bibitem{Lee12} Lee, Seung Hoon, Muhan Choi, Teun-Teun Kim, Seungwoo Lee, Ming Liu, Xiaobo Yin, Hong Kyw Choi et al. ``Switching terahertz waves with gate-controlled active graphene metamaterials." Nature materials 11, no. 11 (2012): 936-941.

\bibitem{Otsuji12} Otsuji, T., SA Boubanga Tombet, A. Satou, H. Fukidome, M. Suemitsu, E. Sano, V. Popov, M. Ryzhii, and V. Ryzhii. ``Graphene-based devices in terahertz science and technology." Journal of Physics D: Applied Physics 45, no. 30 (2012): 303001.

\bibitem{Vicarelli12} Vicarelli, L., M. S. Vitiello, D. Coquillat, A. Lombardo, A. C. Ferrari, W. Knap, M. Polini, V. Pellegrini, and A. Tredicucci. ``Graphene field-effect transistors as room-temperature terahertz detectors." Nature materials 11, no. 10 (2012): 865-871.

\bibitem{Bao12} Bao, Qiaoliang, and Kian Ping Loh. ``Graphene photonics, plasmonics, and broadband optoelectronic devices." ACS nano 6, no. 5 (2012): 3677-3694.

\bibitem{Bonaccorso12} Bonaccorso, Francesco, Z. Sun, T. Hasan, and A. C. Ferrari. ``Graphene photonics and optoelectronics." Nature Photonics 4, no. 9 (2010): 611-622.
`
\bibitem{Hum13} Hum, S., and Julien Perruisseau-Carrier. ``Reconfigurable reflectarrays and array lenses for dynamic antenna beam control: a review." (2013): 1-1.

\bibitem{Yu11} Yu, Nanfang, Patrice Genevet, Mikhail A. Kats, Francesco Aieta, Jean-Philippe Tetienne, Federico Capasso, and Zeno Gaburro. ``Light propagation with phase discontinuities: generalized laws of reflection and refraction." Science 334, no. 6054 (2011): 333-337.

\bibitem{Ni12} Ni, Xingjie, Naresh K. Emani, Alexander V. Kildishev, Alexandra Boltasseva, and Vladimir M. Shalaev. ``Broadband light bending with plasmonic nanoantennas." Science 335, no. 6067 (2012): 427-427.

\bibitem{Nikitin14} Nikitin, A. Yu, T. Low, and L. Martin-Moreno. ``Anomalous reflection phase of graphene plasmons and its influence on resonators." Physical Review B 90, 041407R (2014): 041407.

\bibitem{Efetov10} Efetov, Dmitri K., and Philip Kim. ``Controlling electron-phonon interactions in graphene at ultrahigh carrier densities." Physical review letters 105, no. 25 (2010): 256805.

\bibitem{Wunsch06} Wunsch, B., T. Stauber, F. Sols, and F. Guinea. ``Dynamical polarization of graphene at finite doping." New Journal of Physics 8, no. 12 (2006): 318.

\bibitem{Hwang07} Hwang, E. H., and S. Das Sarma. ``Dielectric function, screening, and plasmons in two-dimensional graphene." Physical Review B 75, no. 20 (2007): 205418.

\bibitem{Palik98} Palik, Edward D., ed. Handbook of optical constants of solids. Vol. 3. Academic press, 1998.

\bibitem{Carrasco13} Carrasco, Eduardo, Michele Tamagnone, and Julien Perruisseau-Carrier. ``Tunable graphene reflective cells for THz reflectarrays and generalized law of reflection." Applied Physics Letters 102, no. 10 (2013): 104103.

\bibitem{Kamoda11} H. Kamoda, T. Iwasaki, J. Tsumochi, T. Kuki, O. Hashimoto, ``60-GHz Electronically Reconfigurable Large Reflectarray Using Single-Bit Phase Shifters", IEEE Trans. on Antennas and Propagation, Vol. 59, No. 7, pp. 2524-2531 (2011)

\bibitem{Woessner14} Woessner, Achim, Mark B. Lundeberg, Yuanda Gao, Alessandro Principi, Pablo Alonso-González, Matteo Carrega, Kenji Watanabe et al. "Highly confined low-loss plasmons in graphene-boron nitride heterostructures." arXiv preprint arXiv:1409.5674 (2014).

\end{thebibliography}

\end{document}